\documentclass[aps,superscriptaddress,showpacs,pre,10pt,nofootinbib,twocolumn]{revtex4-2}
\usepackage{bm} 
\usepackage{graphicx} 
\usepackage{amsmath}
\newcommand\norm[1]{\left\lVert#1\right\rVert}
\usepackage{amsthm}
\usepackage{amssymb} 
\usepackage{epstopdf} 
\usepackage{comment}
\usepackage{amsfonts}
\usepackage{epsfig}
\usepackage{color} 
\usepackage[dvipsnames]{xcolor}
\usepackage{subfigure}
\usepackage[english]{babel}
\usepackage[bb=boondox]{mathalfa}
\usepackage{mathtools}
\usepackage{dsfont}
\usepackage{tikz}
\usetikzlibrary{arrows.meta,
                chains,
                positioning,
                shapes.geometric,
                shadows,shapes,positioning,trees,
                quantikz2, angles, quotes,
                calc
                }
\tikzset{block/.style={draw,minimum height=2cm,minimum width=2cm}}

\usepackage{hyperref}
\usepackage[normalem]{ulem}
\hypersetup{
    colorlinks=true,
    linkcolor=cyan,
    citecolor=magenta,
    filecolor=magenta,
    urlcolor=cyan,
    runcolor=cyan
}

\newcommand {\be}{\begin{equation}}
\newcommand {\ee}{\end{equation}}

\newcommand{\ba}{\begin{eqnarray}}
\newcommand{\ea}{\end{eqnarray}}

\newcommand{\ignore}[1]{}

\usepackage[margin=1in,nofoot]{geometry}

\renewcommand{\Re}{\operatorname{Re}}

\newcommand{\beq}{\begin{equation}}
\newcommand{\eeq}{\end{equation}}
\newcommand{\beqnn}{\begin{equation*}}
\newcommand{\eeqnn}{\end{equation*}}
\newcommand{\bea}{\begin{eqnarray}}
\newcommand{\eea}{\end{eqnarray}}
\newcommand{\beann}{\begin{eqnarray*}}
\newcommand{\eeann}{\end{eqnarray*}}
\newcommand{\bes} {\begin{subequations}}
\newcommand{\ees} {\end{subequations}}

\begin{document}
\raggedbottom 
\title{Permutation Matrix Representation for Quantum Simulation: \\ Comparative Resource Analysis}
\author{Hriday Sabharwal}
\email{hsabharw@usc.edu}
\affiliation{Department of Physics and Astronomy, University of Southern California, Los Angeles, California 90089, USA}
\author{Itay Hen}
\email{itayhen@isi.edu}
\affiliation{Department of Physics and Astronomy, University of Southern California, Los Angeles, California 90089, USA}
\affiliation{Information Sciences Institute, University of Southern California, Marina del Rey, California 90292, USA}

\date{\today}

\begin{abstract}
\noindent 
We present a comparative study of the permutation matrix representation (PMR) method for Hamiltonian simulation alongside other leading quantum algorithms. Our analysis focuses on resource costs for simulating both time-independent and time-dependent Hamiltonians. For the time-independent case, we benchmark PMR against quantum signal processing (QSP) and qubitization, using the Rydberg interaction Hamiltonian as a representative example. For the time-dependent case, we compare the time-dependent extension of PMR with the quantum highly oscillatory protocol (qHOP), applied to a Floquet-driven transverse field Ising model in arbitrary spatial dimensions. In both regimes, we find that PMR offers complementary advantages in resource requirements and exhibits favorable scaling with certain system parameters, suggesting that it may provide practical benefits on resource-constrained quantum hardware.
\end{abstract}

\maketitle

\section{Introduction}

The vision of quantum computing was first motivated by the challenge of simulating quantum systems, for which classical methods generically incur computational costs that scale prohibitively with system size. As the field has evolved, Hamiltonian simulation has emerged as a cornerstone problem, central to advancing our understanding of quantum dynamics in physics, chemistry, and materials science.

The quest for more efficient simulation methods has led to the development of a rich hierarchy of quantum algorithms, each improving upon its predecessors in aspects such as resource overhead, error scaling, and query complexity. These considerations have become especially salient in the noisy, resource-limited era of current quantum hardware, where practical feasibility often hinges on minimizing such costs.

Among the earliest and most widely used techniques is the Suzuki--Trotter product formula, commonly known as Trotterization~\cite{Childs_2021}. While conceptually simple and broadly applicable, Trotterization incurs gate counts that scale superoptimally with the target error tolerance. More advanced approaches include the Linear Combination of Unitaries (LCU) method~\cite{Berry_2015}, which leverages the prepare--select framework to implement a truncated Taylor expansion of the exponential, and quantum signal processing (QSP), or qubitization~\cite{Low_2019}, which employs a sophisticated interplay between quantum circuits and classical preprocessing to approximate the exponential via a truncated Jacobi--Anger expansion. These latter algorithms achieve asymptotically near-optimal scaling but remain challenging to implement on near-term platforms.

Recent progress on the Permutation Matrix Representation (PMR) algorithm for Hamiltonian simulation~\cite{Kalev_2021,Kalev_2025} offers an alternative whose resource requirements can be parametrically favorable in regimes where diagonal contributions dominate the Hamiltonian norm. The key feature of PMR lies in its use of a permutation matrix expansion, which decouples diagonal and off-diagonal contributions to the Hamiltonian. When combined with a truncated Taylor expansion, this structure can eliminate dependence on parameters governing the diagonal sector, potentially yielding improved efficiency for systems where diagonal terms dominate the norm.

The PMR framework extends naturally to time-dependent Hamiltonians. In this setting, the evolution operator can be expressed as a sum of unitaries using divided differences of the exponential, leading to an integral-free Dyson-series expansion. This representation facilitates an efficient truncated-series approximation implemented through the LCU paradigm~\cite{Chen_2021}.

In this work, we conduct a systematic comparative study of PMR alongside leading near-optimal algorithms, focusing on models of practical interest. For the time-independent case, we benchmark PMR against qubitization using the Rydberg interaction Hamiltonian. For the time-dependent case, we compare the PMR algorithm with the recently proposed quantum highly oscillatory protocol (qHOP)~\cite{An_2022}, using a Floquet-driven transverse field Ising model in $d$ spatial dimensions as a testbed.

The remainder of the paper is organized as follows. Section~\ref{sec:timeindep} provides an overview of qubitization and PMR as applied to the Rydberg atom Hamiltonian, along with a comparative analysis of the resource costs. Section~\ref{sec:timedep} focuses on the time-dependent case, briefly reviewing qHOP and PMR, and providing a similar comparative analysis between resource costs as applied to the Floquet-driven transverse field Ising model. We conclude in Section~\ref{sec:summary} with a summary and discussion of broader implications.

\section{Time-independent Hamiltonian simulation: PMR vs Qubitization\label{sec:timeindep}}

In this section, we analyze the asymptotic resource scaling of the PMR algorithm when applied to the dynamical simulation of a model of physical relevance, specifically the Rydberg atom Hamiltonian. For context, we benchmark its performance against qubitization, a state-of-the-art quantum simulation framework for time-independent Hamiltonians.

Our analysis demonstrates that while both PMR and qubitization are capable of reproducing the target dynamics with high fidelity, each approach has distinct advantages depending on the system parameters. To establish this comparison rigorously, we begin with a concise overview of the two methods. We outline the fundamental principles underlying each algorithm, with a particular emphasis on the computational structures and approximation strategies that ultimately determine their resource requirements.

\subsection{Time-independent PMR}
The PMR approach separates the diagonal and non-diagonal evolutions by using divided differences. The definitions and results presented in this subsection follow Ref.~\cite{Kalev_2021}.

The PMR form of the Hamiltonian is written as:
\begin{equation}\label{eq:1}
H = D_0 + \sum_{i=1}^M D_i P_i\ ,
\end{equation}
where the $D_i$ are diagonal operators and the $P_i$ are permutation operators. This decomposition enables an expansion of the evolution operator in terms of divided differences~\cite{Kalev_2021}. The total time is divided into $r$ intervals of duration $\Delta t = t/r$ to optimize resources~\cite{Kalev_2021}. The expansion takes the form:
\begin{equation}\label{eq:2}
\begin{aligned}
U&:= e^{-iH\Delta t} = \sum_{n=0}^\infty \frac{(-i\Delta t)^n}{n!}H^n \\
&= \sum_{n=0}^\infty \frac{(-i\Delta t)^n}{n!}\left(\sum_{i=0}^M D_i P_i\right)^n\ .
\end{aligned}
\end{equation}
After some algebra~\cite{Albash_2017,Gupta_2020}, this expansion can be expressed as:
\begin{equation}\label{eq:3}
U=U_{\text{od}}e^{-i\Delta tD_0}\ ,
\end{equation}
where
\begin{equation}
    U_{\text{od}}:=\sum_z\sum_{q=0}^\infty\sum_{\mathbf{i}_q} e^{-i\Delta t[\Delta E_z,...,\Delta E_{z_q}]}d_{\mathbf{i}_q}P_{\mathbf{i}_q}\ket{z}\bra{z}\ ,
\end{equation}
and where $\mathbf{i}_q = (i_1, i_2, \ldots, i_q)$ is a tuple of $q$ indices each ranging from 1 to $M$ and $P_{\mathbf{i}_q} := P_{i_q}\cdots P_{i_2} P_{i_1}$ is an ordered product of off-diagonal permutation operators. Here $e^{-i\Delta t[\Delta E_z,\ldots,\Delta E_{z_q}]}$ is the divided difference exponential, where $\Delta E_{z_j}=E_{z_j}-E_z$, $E_{z_j} = \langle z_j | D_0 | z_j \rangle$, $|z_j\rangle = P_{i_j}\cdots P_{i_1}|z\rangle$ and the convention $z_0 := z$ is chosen. $d_{\mathbf{i}_q} = d_{i_q}\cdots d_{i_2}d_{i_1}$ where $d_{i_j} = \langle z_j | D_{i_j}| z_j \rangle$.

Thus we can see that the PMR expansion effectively decouples the evolutions due to the diagonal and the off-diagonal parts of the Hamiltonian. The diagonal evolution admits a straightforward circuit implementation, whose cost we quantify in the following subsection. In order to simulate $U_{\text{od}}$, we will employ the use of LCU~\cite{Berry_2015}. In order to use LCU, we will need to write $U_{\text{od}}$ as a sum of unitaries.

This can be done by noticing that the complex number 
\begin{equation}
    \beta_{\mathbf{i}_q}^{(z)}:=\frac{q!}{\Gamma_{\mathbf{i}_q}(\Delta t)^q}e^{-i\Delta t[\Delta E_z,...,\Delta E_{z_q}]}d_{\mathbf{i}_q}
\end{equation}
has norm less than 1, where we have defined $\Gamma_{\mathbf{i}_q}=\prod_{j=1}^q\Gamma_{i_j}$ and $\Gamma_i\geq\norm{D_i}_{\max}$. Hence this number can be written as a sum of phases:
\begin{equation}
\begin{aligned}
    \beta_{\mathbf{i}_q}^{(z)}&=\cos\phi_{\mathbf{i}_q}^{(z)}e^{i\chi_{\mathbf{i}_q}^{(z)}}\\&=\frac{1}{2}\left(e^{i(\chi_{\mathbf{i}_q}^{(z)}+\phi_{\mathbf{i}_q}^{(z)})}+e^{i(\chi_{\mathbf{i}_q}^{(z)}-\phi_{\mathbf{i}_q}^{(z)})}\right)\ .
    \end{aligned}
\end{equation}
Thus we can rewrite $U_{\text{od}}$ as:
\begin{equation}
    U_{\text{od}}=\sum_{k=0,1}\sum_{q=0}^\infty\sum_{\mathbf{i}_q}\frac{\Gamma_{\mathbf{i}_q}(\Delta t)^q}{2q!}U_{\mathbf{i}_q}^{(k)}\ ,
\end{equation}
where
\begin{equation}
    U_{\mathbf{i}_q}^{(k)}=P_{\mathbf{i}_q}\Phi_{\mathbf{i}_q}^{(k)}\ ,
\end{equation}
and $\Phi_{\mathbf{i}_q}^{(k)}=\sum_ze^{i(\chi_{\mathbf{i}_q}^{(z)}+(-1)^k\phi_{\mathbf{i}_q}^{(z)})}\ket{z}\bra{z}$ is a diagonal unitary. 

In order to implement the LCU, first the infinite sum is truncated to order $Q$, and we define
\begin{equation}
    \tilde{U}_{\text{od}}:=\sum_{k=0,1}\sum_{q=0}^Q\sum_{\mathbf{i}_q}\frac{\Gamma_{\mathbf{i}_q}(\Delta t)^q}{2q!}U_{\mathbf{i}_q}^{(k)}\ .
\end{equation}
We define the off-diagonal norm $\Gamma:=\sum_{i=1}^M\Gamma_i$, which will be used later in the resource cost analysis.

After this step, the algorithm follows the standard LCU procedure~\cite{Berry_2015} to implement this approximation. Finally, $\tilde{U}:=\tilde{U}_{\text{od}}e^{-i\Delta tD_0}$ must be applied $r$ times to obtain the final approximation.

\subsection{Qubitization}

The qubitization approach implements an approximation to the evolution operator through a two-stage process. First, the Hamiltonian is block encoded. Then, quantum signal processing is used to implement an approximate exponential function on the block encoding. Projecting back to the system subspace yields the desired expansion. The definitions and results presented in this subsection are based on Ref.~\cite{Low_2019}.

Block encoding uses a carefully chosen ancilla state $|G\rangle_a$ and its orthogonal subspace to expand the Hilbert space and encode the Hamiltonian in the $|G\rangle_a$, $|G^\perp\rangle_a$ basis as:
\begin{equation}\label{eq:9}
U = \begin{pmatrix}
H & \cdot \\
\cdot & \cdot
\end{pmatrix}
\end{equation}
\begin{equation}\label{eq:10}
H = (\langle G|_a \otimes I_s) U (|G\rangle_a \otimes I_s)\ .
\end{equation}
Let us denote $|G\rangle_a$ simply as $|G\rangle$. The operator $U$ is not ideal for repeated applications because the subspace $\mathcal{H}_\lambda := \text{span}\{|G\rangle|\lambda\rangle, U|G\rangle|\lambda\rangle\}$ for each eigenstate is not invariant under $U$. This problem can be addressed using the iterate $W$. Qubitization provides a method to construct this iterate $W$ with minimal overhead~\cite{Low_2019}, involving a reflection about $|G\rangle$:
\begin{equation}\label{eq:11}
W = ((2|G\rangle\langle G| - I)_a \otimes I_s) SU\ ,
\end{equation}
where $S$ can be constructed using control-$U$ and an extra ancilla. The detailed construction of this iterate can be found in Ref.~\cite{Low_2019}. Successive applications of $W$ will produce Chebyshev polynomials. By introducing the phased iterate $W_\phi$ ($V_\phi$ in the case of single ancilla QSP) and by successively implementing them with different phases, the polynomial that is constructed can be controlled by choosing the phases appropriately. The number of iterates needed is determined by the required accuracy of the approximation, which in turn depends on the truncation of the polynomial series that approximates the target unitary:
\begin{equation}\label{eq:12}
V_{\vec{\phi}} = \prod_{\substack{k \text{ odd} \\ k \geq 1}}^{Q/2} V^\dagger_{\phi_{k+1}+\pi} V_{\phi_k}\ .
\end{equation}
Similar to previous approaches~\cite{Low_2017,Berry_2015b}, in order to implement the exponential evolution operator, the Jacobi-Anger series is chosen~\cite{Abramowitz:1964:HMF}.

\subsection{Resource cost analysis for the Rydberg atom Hamiltonian}

The Rydberg interaction Hamiltonian~\cite{Lukin_2001,Saffman_2010,Jaksch_2000,Bernien_2017,Dudin_2012} represents the dynamics of a system of neutral atoms, excited to their Rydberg states through coherent driving by a laser. The driving gives rise to transitions between ground and excited Rydberg states, and the excited atoms allow for van der Waals interactions among themselves. This can lead to the Rydberg blockade phenomenon, wherein one atom in the Rydberg state can prevent the excitation of nearby atoms.

The $N$-atom Rydberg Hamiltonian is generally given as:
\begin{equation}\label{eq:rh}
H_{\text{Ryd}} = \frac{1}{2}\sum_{i=1}^N (\Omega_i X_i - \delta Z_i) + \sum_{i<j}\frac{C_6}{|\textbf{r}_i-\textbf{r}_j|^6}n_i n_j\ .
\end{equation}
Here, the $\Omega_i$ are the Rabi frequencies, describing the coupling strength between ground and excited states of the $i$-th atom. The terms $\frac{\delta}{2}Z_i$ describe the detunings of the driving laser from the excited Rydberg state, and $n_i = \frac{1}{2}(I+Z_i)$ are the number operators whose eigenstates are the ground and excited states. The interaction term $\frac{C_6}{|\textbf{r}_i-\textbf{r}_j|^6}n_i n_j$ represents the van der Waals interaction between two atoms in their Rydberg states, with strength depending on their spatial separation.

We will assume that the atoms are equally separated by some distance $R$, and for the sake of brevity in later sections, we will define $C_6':=\frac{C_6}{R^6}$.

\subsubsection{With qubitization}\label{sec:qubitization}

We write the Hamiltonian in LCU block encoding~\cite{Low_2019} by decomposing it entirely in terms of Pauli strings. For simplicity, we present the detailed computation in Appendix~\ref{A} and summarize the results here. The Hamiltonian can be written as:
\begin{equation}\label{eq:14}
H_{\text{Ryd}} := \sum_{i=1}^M \alpha_i P_i\ ,
\end{equation}
where $\alpha_i \geq 0$ and the $P_i$ are Pauli strings which include any negative signs, and where $\alpha := \sum_{i=0}^M |\alpha_i|$. Here $M = \mathcal{O}(N^2)$ because of the $\sum_{i<j}$ term. Following the LCU framework~\cite{Berry_2015}, we define the operators:
\begin{equation}\label{eq:15}
\text{PREP}|0\rangle := \frac{1}{\sqrt{\alpha}}\sum_{i=1}^M \sqrt{\alpha_i}|i\rangle
\end{equation}
\begin{equation}\label{eq:16}
\text{SEL} := \sum_{i=1}^M |i\rangle\langle i| \otimes P_i\ ,
\end{equation}
where PREP acts on a set of ancilla qubits and SEL acts on both the ancilla and the system qubits. Then we can construct the corresponding block encoding as:
\begin{equation}\label{eq:17}
U := (\text{PREP}^\dagger_a \otimes I_s) \text{ SEL } (\text{PREP}_a \otimes I_s)
\end{equation}
\begin{equation}\label{eq:18}
|G\rangle := |0\rangle_a\ ,
\end{equation}
such that:
\begin{equation}\label{eq:19}
\langle G|U|G\rangle = \frac{1}{\alpha}H_{\text{Ryd}}\ .
\end{equation}
It can be shown that:
\begin{equation}\label{eq:20}
\alpha := \sum_{i=1}^M |\alpha_i| \approx \mathcal{O}\left(N\left(\frac{\Omega}{2} + \frac{\delta}{2} + C_6'\right)\right)\ ,
\end{equation}
where $\Omega$ is the average of all the $\Omega_i$'s. Although the double sum naïvely suggests $\mathcal{O}(N^2)$ scaling, the rapid decay of $C_{ij}$ with $|i-j|$ ensures that $\sum_{i<j} C_{ij}$ is in fact at most $\mathcal{O}(N)$.

Next, we construct the iterate $W$ as described in Eq.~(\ref{eq:11}). The cost depends on that of constructing $U$, which in turn depends on the cost of the SEL and PREP operators. Following Ref.~\cite{Berry_2015}, this cost is:
\begin{equation}
    \text{PREP}=\mathcal{O}\left(M\frac{\log(\alpha t/\epsilon)}{\log\log(\alpha t/\epsilon)}\right)\ ,
\end{equation}
and 
\begin{equation}
    \text{SEL}=\mathcal{O}\left(M(N+\log M)\frac{\log(\alpha t/\epsilon)}{\log\log(\alpha t/\epsilon)}\right)\ .
\end{equation}
We show in the next subsection that the $\log/\log\log$ dependence scales very slowly asymptotically, and may be omitted from the final cost.

The next step is to create the phased iterate $V_\phi$; however, this step does not involve implementing any processes that exceed the gate cost. The next step, which involves creating the sequence of phased iterations, does increase the cost.

Next, as also described by Eq.~(\ref{eq:12}), we need $Q$ repetitions of $V_\phi$, where $Q$ corresponds to the truncation in the Jacobi-Anger expansion. To achieve an error of $\mathcal{O}(\epsilon)$ in the expansion, we have~\cite{Low_2019}:
\begin{equation}\label{eq:21}
Q = \mathcal{O}(\alpha t + \log(1/\epsilon))\ .
\end{equation}
Hence, the total gate cost of implementing the algorithm becomes:
\begin{equation}\label{eq:22}
\begin{aligned}
\tilde{U} &= \mathcal{O}(QW) \\
&\approx \mathcal{O}\left(\left(\alpha t + \log(1/\epsilon)\right)(MN)\right) \\
&\approx \mathcal{O}\left(N\left(\frac{\Omega}{2}+\frac{\delta}{2}+C_6'\right)t + \log(1/\epsilon)\right)(N^3) \\
&= \mathcal{O}\left(N^4 t(\Omega + \delta + C_6') + N^3\log(1/\epsilon)\right)\ .
\end{aligned}
\end{equation}
The qubit cost of the qubitization approach stems mainly from the block-encoding i.e. from the $\text{PREP}_a$ operation, requiring $\mathcal{O}(\log M)\approx\mathcal{O}(\log N)$ ancilla qubits. There will be two additional ancilla qubits required for qubitization, but they may be omitted from the overall qubit cost.

\subsubsection{With PMR}

We begin by writing the Hamiltonian in the PMR form. This is straightforward since the only non-diagonal term in the Hamiltonian is the Pauli $X$ operator. The Hamiltonian can be written as:
\begin{equation}\label{eq:23}
H_{\text{Ryd}} = D_0 + \sum_{i=1}^M D_i P_i\ ,
\end{equation}
where
\begin{equation}\label{eq:24}
D_0 = -\frac{\delta}{2}\sum_{i=1}^N Z_i + \sum_{i<j}\frac{C_6}{|\textbf{r}_i-\textbf{r}_j|^6}n_i n_j\ ,
\end{equation}
and $P_i = X_i$, $D_i = \frac{\Omega_i}{2}\cdot I$. From this decomposition, we can see that the off-diagonal norm is $\Gamma = \frac{1}{2}\sum_i \Omega_i = \frac{N\Omega}{2}$, and that the number of terms in the decomposition is $M = N$.

The main resource cost in the PMR algorithm stems from the implementation of the controlled unitaries in the LCU. Following the analysis in~\cite{Kalev_2021}, the gate cost to implement the short time approximation $\tilde{U}$ scales as $\mathcal{O}(C_{D_0}+Q^2+QM(C_{\Delta D_0}+k_{\text{od}}+\log M))$ with qubit cost scaling as $\mathcal{O}(Q\log M)$. Here, $C_{D_0}$ is the cost of calculating a digonal energy (a single $D_0$ matrix element), $C_{\Delta D_0}$ is the cost of calculating the change to a diagonal energy due to the action of a $P_i$, and $k_{\text{od}}$ is the upper bound on the locality of $P_i$.

To calculate $C_{\Delta D_0}$, note that $P_i=X_i$, so acting a permutation on $D_0$ will flip the sign of only one term in the first sum and $N+2$ terms in the second sum. Thus $C_{\Delta D_0}=\mathcal{O}(N)$. For $C_{D_0}$, the number of terms in the expansion of $D_0$ is $\mathcal{O}(N^2)$, and there are at most 2 $Z$ operators in the summand. Thus $C_{D_0}=\mathcal{O}(2N^2)\equiv\mathcal{O}(N^2)$.

Each $P_i$ consists of a single $X$ operator, so $k_{\text{od}}=1$. Next, setting $r=\lceil t\Gamma/\ln(2)\rceil$ ensures that the sum of coefficients in the LCU is approximately equal to 2, which is required for oblivious amplitude amplification~\cite{Berry_2015}. Furthermore, setting
\begin{equation}
    Q=\mathcal{O}\left(\frac{\log(t\Gamma/\epsilon)}{\log\log(t\Gamma/\epsilon)}\right)
\end{equation}
ensures that the final short time approximation $\tilde{U}$ is within error $\epsilon/r$ of the actual short time evolution $U$~\cite{Berry_2015}~\cite{Kalev_2021}.

It can be verified that:
\begin{equation}\label{eq:27}
\frac{\log x}{\log\log x} < x^s \quad \text{as } x \to \infty, \quad \forall s > 0\,.
\end{equation}
This implies that the $\log/\log\log$  dependence is sub-polynomial. Up to polylogarithmic corrections, we may omit the explicit $Q$ dependence in the simulation cost, giving us:
\begin{equation}\label{eq:pmrcost}
\mathcal{O}\left(tN^3\Omega + t\Omega N^2\log N\right)\ ,
\end{equation}
along with an approximate qubit cost $\mathcal{O}(\log N)$.

We also highlight a method to approximate the diagonal evolution and the energy differences, which further improves the PMR gate cost by allowing for some error in their implementation. We describe this method and the involved computations in Appendix~\ref{C} and~\ref{D}.

Another important note should be made regarding the improved implementation of the PMR algorithm presented in~\cite{Kalev_2025}. The algorithm presented in~\cite{Kalev_2021} uses binary encoding for the state preparation sub-routine, whereas~\cite{Kalev_2025} leverages unary encoding for state preparation, along with the help of another approximation involving successive applications of the Leibniz rule for divided differences. These new aspects allow for a simpler implementation of the select unitary operations requiring only the use of CNOT gates, at the cost of additional qubits required for the unary encoding.

\subsubsection{Comparison}

\begin{table*}[tb]
\begin{center}
\begin{tabular}{| c | c | c |} 
 \hline \hline
 Algorithm & Gate Cost & Qubit cost \\ [0.5ex]
 \hline
 Qubitization & $\mathcal{O}\left(N^4 t(\Omega + \delta + C_6') + N^3\log(1/\epsilon)\right)$ & $\mathcal{O}(\log N)$\\ [1ex]
 \hline
 PMR & $\mathcal{O}\left(N^3t\Omega + t\Omega N^2\log N\right)$ & $\mathcal{O}(\log N)$\\ [1ex]
 \hline
 PMR approximation & $\mathcal{O}\left(tN^2\Omega\left(\frac{tC_6'}{\epsilon }\right)^{1/5}+t\Omega N^2\log N\right)$ & $\mathcal{O}(\log N)$\\ [1ex]
 \hline
\end{tabular}
\caption{\label{tab:timeindep}\textbf{Comparison of algorithm costs for time-independent simulation.} While qubitization achieves better gate cost scaling than many approaches, the PMR approach offers a superior scaling in terms of various system parameters. We can further utilize the Hamiltonian structure to implement an approximation of PMR, which has even better scaling in terms of $N$. The qubit cost for all the approaches is identical.}
\end{center}
\end{table*}

The two approaches exhibit different scaling behaviors, as summarized in Table~\ref{tab:timeindep}. A notable distinction is that PMR removes explicit dependence on the magnitude of diagonal terms (though complexity still depends on the cost of evaluating diagonal energies) whereas qubitization's complexity depends on all three parameters $\Omega$, $\delta$, and $C_6$. This means that the PMR simulation cost remains unchanged regardless of the magnitudes of $\delta$ or $C_6$, while qubitization's cost scales linearly with these parameters. This feature arises from PMR's decoupling of diagonal and off-diagonal evolution: since $\delta$ and $C_6$ belong to the diagonal part, they do not contribute to the resource scaling.

The scaling with respect to $N$ also differs between the two methods. For systems with a large number of qubits, PMR's $\mathcal{O}(N^3)$ scaling may be advantageous compared to qubitization's $\mathcal{O}(N^4)$ scaling. This difference can be traced to the LCU block encoding used in qubitization, which must account for $\mathcal{O}(N^2)$ Pauli strings when expanding the interaction term. In contrast, PMR treats the interaction term as part of the diagonal evolution, with only the off-diagonal terms contributing at $\mathcal{O}(N)$. We note that the $\mathcal{O}(N^3)$ scaling in qubitization may be improved with more efficient block encoding strategies. The dependence on $N$ arising from the norm (or the off-diagonal norm for PMR) is comparable for both algorithms. Apart from the gate costs, the qubit cost for both the algorithms scales as $\mathcal{O}(\log N)$.

We also include the method which approximates diagonal evolution and energy differences in Table~\ref{tab:timeindep}. Similar to $\alpha$ of Section \ref{sec:qubitization}, we take advantage of the Hamiltonian structure by observing that the coefficients of the double sum in $D_0$ fall off very quickly. This suggests that we can cut off this sum at a carefully chosen order, allowing us to implement the required operations at a lower cost, with some incurred error. The main advantage of this approximation lies in the further reduced scaling with $N$, at the cost of an additional dependence on other parameters that scales only as a small polynomial.

\section{Time-dependent Hamiltonian simulation: PMR vs qHOP\label{sec:timedep}}

Next, we turn to consider the time-dependent case. Here we compare the PMR method with qHOP, which represents a state-of-the-art approach that performs well in challenging regimes, particularly when the Hamiltonian has a large norm or exhibits rapid time variation.
To understand the relative performance characteristics, we analyze both algorithms on the Floquet-driven transverse field Ising model, examining their dependence on the various model parameters.
As before, we first provide an overview of both methods, focusing on the approximations that contribute to the resource cost.

\subsection{Time-dependent PMR}

The time-dependent version of the PMR algorithm follows the same conceptual framework as the time-independent algorithm. The algorithm approximates the time-ordered exponential as a sum of unitaries using divided differences. This approximation is then implemented via LCU. The definitions and results in this subsection follow Ref.~\cite{Chen_2021}.

First, the Hamiltonian is expanded in a manner similar to the time-independent case, but now we include the time-dependent parts as time-dependent diagonal matrices:
\begin{equation}\label{eq:28}
H(t) = \sum_{i=0}^M D_i(t) P_i\ ,
\end{equation}
where the 0-th index represents the diagonal part of the Hamiltonian. The time-dependent diagonal matrices are further decomposed into a finite sum of exponential functions:
\begin{equation}\label{eq:29}
D_i(t) = \sum_{k=1}^{K_i} \exp\left(\Lambda_i^{(k)} t\right) D_i^{(k)}\ ,
\end{equation}
where $\Lambda_i^{(k)}$ and $D_i^{(k)}$ are complex diagonal matrices with diagonal elements:
\begin{equation}\label{eq:30}
\lambda_{i,z}^{(k)} = \langle z|\Lambda_i^{(k)}|z\rangle
\end{equation}
\begin{equation}\label{eq:31}
d_{i,z}^{(k)} = \langle z|D_i^{(k)}|z\rangle\ .
\end{equation}
We set $K_i = K$ for all $i$, as we are free to set any extra terms in the sum to zero.

We then proceed to insert this expansion in the time evolution operator:
\begin{equation}\label{eq:32}
U(t) = \mathcal{T} \exp\left(-i\int_0^t H(t')dt'\right)\ .
\end{equation}
After inserting the expansion~(\ref{eq:28}) in the above expression and performing the algebraic manipulations detailed in~\cite{Chen_2021}, it can be shown that the action of the evolution operator on a basis vector $|z\rangle$ is:
\begin{equation}\label{eq:33}
\begin{aligned}
U(t)|z\rangle &= \sum_{q=0}^\infty \sum_{\mathbf{i}_q}\sum_{\mathbf{k}_q} (-i)^q \int_0^t d\tau_q \cdots \int_0^{\tau_2} d\tau_1 \\
&\quad \times \exp\left(\lambda_{\mathbf{i}_q,z_{\mathbf{i}_q}}^{(\mathbf{k}_q)}\tau_q + \cdots + \lambda_{\mathbf{i}_1,z_{\mathbf{i}_1}}^{(\mathbf{k}_1)}\tau_1\right) \\
&\quad \times d_{\mathbf{i}_q,z}^{(\mathbf{k}_q)} P_{\mathbf{i}_q}|z\rangle\ ,
\end{aligned}
\end{equation}
where $\mathbf{i}_q = \{i_q, \ldots, i_1\}$ and $\mathbf{k}_q = \{k_q, \ldots, k_1\}$ denote multi-indices, $\{z_{\mathbf{i}_j}\} = P_{i_j}\cdots P_{i_1}|z\rangle$ (where $1 \leq j \leq q$), $P_{\mathbf{i}_q} = P_{i_q}\cdots P_{i_1}|z\rangle$ and $d_{\mathbf{i}_q,z}^{(\mathbf{k}_q)} = d_{i_q,z_{i_q}}^{(k_q)}\cdots d_{i_1,z_{i_1}}^{(k_1)}$.

The expansion can be written in an integral-free form using the identity~\cite{Kalev_2021b}:
\begin{equation}\label{eq:34}
\int_0^1 ds_q \cdots \int_0^{s_2} ds_1 \, e^{(\lambda_1 s_1 + \cdots + \lambda_q s_q)} = e^{[x_1,\cdots,x_q,0]}\ ,
\end{equation}
where $x_j = \sum_{l=j}^q \lambda_l$ and $e^{[x_1,\cdots,x_q,0]}$ is the divided difference of the exponential function.

Using this, and by completing the basis, we obtain:
\begin{equation}\label{eq:35}
\begin{aligned}
U(t) &= \sum_z U(t)|z\rangle\langle z| \\
&= \sum_{z,\mathbf{i}_q,\mathbf{k}_q} \sum_{q=0}^\infty (-i)^q e^{t[x_1,\cdots x_q,0]} d_{\mathbf{i}_q,z}^{(\mathbf{k}_q)} P_{\mathbf{i}_q}|z\rangle\langle z|\ .
\end{aligned}
\end{equation}

Now the algorithm divides the total time $T$ into $r$ pieces and uses the LCU routine to implement Eq.~(\ref{eq:35}) for each small time, by truncating the sum at some order $Q$~\cite{Chen_2021}.

The algorithm can also be modified to incorporate the interaction picture scheme. The overall algorithm remains the same, with the only change appearing in the expansion when, instead of $U(t)$, we use the following evolution operator:
\begin{equation}\label{eq:ui}
U_I(t) = \mathcal{T} \exp\left[-i\int_0^T e^{iH_0 t}V(t)e^{-iH_0 t}\right]\ .
\end{equation}
When the procedure is repeated for the above $U_I(t)$, the diagonal part appears within the expansion. As shown in Ref.~\cite{Chen_2021}, however, the diagonal evolution does not contribute to the complexity and merely appears as phases in the expansion.

\subsection{qHOP}

The qHOP algorithm uses a number of approximations along with the use of QSP~\cite{Gily_n_2019} to implement the evolution operator:
\begin{equation}\label{eq:36}
U(T,0) = \mathcal{T} e^{-i\int_0^T H(s)ds}\ .
\end{equation}
The definitions and results presented in this subsection follow Ref.~\cite{An_2022}. To begin with, the total simulation time $T$ is divided into $L$ time steps of size $h$, i.e., $h = T/L$. The total evolution operator can then be written as a product of small time evolutions:
\begin{equation}\label{eq:37}
\begin{aligned}
U(T,0) &= \prod_{j=0}^{L-1} \mathcal{T} e^{-i\int_{jh}^{(j+1)h} H(s)ds} \\&:= \prod_{j=0}^{L-1} U((j+1)h, jh)\ .
\end{aligned}
\end{equation}
The first approximation drops the time ordering, expressing the short-time evolution as the exponential of the integral:
\begin{equation}\label{eq:38}
U((j+1)h, jh) \approx e^{-i\int_{jh}^{(j+1)h} H(s)ds}\ .
\end{equation}
This corresponds to truncating the Magnus expansion at first order~\cite{Magnus:1954zz,Iserles1999}.

For the next approximation, first-order quadrature is used to approximate the integral:
\begin{equation}\label{eq:39}
\int_{jh}^{(j+1)h} H(s)ds \approx \frac{h}{M}\sum_{k=0}^{M-1} H(jh + (kh/M))\ ,
\end{equation}
where $M$ is the number of nodes for the quadrature. Hence the short time evolution can be written as:
\begin{equation}\label{eq:40}
\begin{aligned}
U((j+1)h, jh) &\approx U_1((j+1)h, jh) \\&:= e^{-i\frac{h}{M}\sum_{k=0}^{M-1} H(jh+(kh/M))}\ .
\end{aligned}
\end{equation}
The final result then becomes:
\begin{equation}\label{eq:41}
U(T,0) \approx \prod_{j=0}^{L-1} U_1((j+1)h, jh)\ .
\end{equation}

Now to implement the operator $U_1$, QSP is utilized. We assume access to an oracle that can produce the Hamiltonian evaluated at different discrete time steps. We refer to this as the HAM-T$_j$ oracle. We also assume two sets of ancilla qubits $a$ and $m$, with the system qubits labeled by $s$. The implementation of this oracle is defined as:
\begin{equation}\label{eq:42}
\begin{aligned}
\langle 0|_a&\text{ HAM-T}_j |0\rangle_a \\&= \frac{1}{\alpha}\sum_{k=0}^{M-1} |k\rangle_m\langle k|_m \otimes H(jh+kh/M)\ ,
\end{aligned}
\end{equation}
where $\|H(t)\| \leq \alpha$ for all $t$. A block encoding can then be defined using Hadamard gates such that the sum in Eq.~(\ref{eq:39}) can be encoded, i.e., define
\begin{equation}\label{eq:43}
U = (I_a \otimes H_m^{\otimes n_m} \otimes I_s) \text{ HAM-T}_j (I_a \otimes H_m^{\otimes n_m} \otimes I_s)
\end{equation}
\begin{equation}\label{eq:44}
|G\rangle = |0\rangle_a \otimes |0\rangle_m\ ,
\end{equation}
then
\begin{equation}\label{eq:45}
\langle G|U|G\rangle = \frac{1}{M\alpha}\sum_{k=0}^{M-1} H(jh + (kh/M))\ ,
\end{equation}
and then QSP~\cite{Gily_n_2019} is used to implement the exponential.

Composing these short-time approximations yields the full evolution operator to the desired accuracy.

The interaction picture scheme can be used to further reduce the cost. If the Hamiltonian is divided into time-independent ($A$) and time-dependent ($B(t)$) parts, this scheme will avoid the possible dependence on the large norm of the time-independent part.

The idea is the same, however instead of $H(t)$ we now deal with $H_I(t)$ defined as:
\begin{equation}\label{eq:46int}
H_I(t) = e^{iAt}B(t)e^{-iAt}\ .
\end{equation}
Using this interaction Hamiltonian in Eq.~(\ref{eq:40}), we get a modified expression for $U_1$:
\begin{equation}\label{eq:47int}
\begin{aligned}
&U_1((j+1)h, jh) = e^{iAjh} \\
& \times \exp\left(-i\frac{h}{M}\left(\sum_{k=0}^{M-1} e^{i\frac{Akh}{M}}B(jh + \frac{kh}{M})e^{-i\frac{Akh}{M}}\right)\right) \\
& \times e^{-iAjh}\ .
\end{aligned}
\end{equation}
So now the block encoding for the HAM-T$_j$ operator in Eq.~(\ref{eq:43}) will also have to be modified accordingly in order to implement the expression in the exponential in Eq.~(\ref{eq:47int}). This block encoding is a bit more involved, but is given in detail in Ref.~\cite{An_2022}.

The interaction picture qHOP algorithm uses two simpler oracles to implement this block encoding, namely the oracle $O_A(s)$, which implements $e^{iAs}$ for any $s$, and the oracle $O_B(j)$, which is simply the HAM-T oracle for $B(t)$.

\subsection{Resource cost analysis for the Floquet-driven transverse field Ising model in $d$-dimensions}

To analyze the time-dependent case, we consider a Floquet-driven transverse field Ising model~\cite{Weitenberg_2021,Sen_2021}. Interest in such models stems primarily from the connection between the zero-temperature critical behavior of quantum spin Ising systems in $d$ dimensions and the critical behavior of the corresponding $d+1$-dimensional classical system~\cite{Sondhi_1997,sachdev2011}, along with their general relevance to quantum magnetism~\cite{RevModPhys.25.353,Haggkvist01092007}.

We define the model on a $d$-dimensional hypercubic lattice with $\tilde{N}$ particles on each axis. The Hamiltonian can be written as:
\begin{equation}\label{eq:46}
H(t) = -J\left(\sum_{\langle i,j\rangle} Z_i Z_j\right) - \zeta\cos(\omega t)\left(\sum_i X_i\right)\ ,
\end{equation}
where $\sum_{\langle i,j\rangle}$ denotes a sum over all nearest neighbors. Here, $J$ is the interaction strength between neighboring spins and $\zeta$ is the driving amplitude of the external field. We assume periodic boundary conditions. This does not affect the analysis, as the edge contributions may be neglected when $\tilde{N}$ becomes large. We also denote the total number of particles as $N := \tilde{N}^d$ for brevity.

We will also need the number of terms in each summation for our analysis. The Floquet term simply involves a sum over all the particles, hence the sum contains $N$ terms. The interaction term contains a sum over all nearest neighbors. We know that each particle is connected to two other particles per axis; hence, the number of nearest neighbors per particle is $2d$ for $d$ dimensions/axes. If we simply count this for all particles, accounting for the double counting, we will get that there are $dN$ total nearest neighbors.

\subsubsection{With qHOP}

We analyze the complexity of the algorithm as applied to the simulation of the above Hamiltonian. The main contributors to the complexity are the construction of the block encoding and the implementation of QSP. We follow corollary 2 of Ref.~\cite{An_2022}, which gives the complexity as
\begin{equation}\label{eq:47}
\begin{aligned}
\mathcal{O}\Bigg(&\min\left\{\frac{\alpha_B^2 T^2}{\epsilon}\log\left(\frac{\alpha_B T}{\epsilon}\right),\right. \\
&\left.\alpha_B T + \frac{\alpha_B^{1/2}(\alpha_{AB} + \beta_B)^{1/2}T^{3/2}}{\epsilon^{1/2}}\right.\\&\times\left.\log\left(\frac{\alpha_B(\alpha_{AB} + \beta_B)T}{\epsilon}\right)\right\} \\
&\times \log\left(\frac{(\alpha_{AB} + \beta_B)T}{\epsilon}\right)\Bigg)
\end{aligned}
\end{equation}
number of uses of the oracles $O_A$ and $O_B$. Here, $\max_{t\in[0,T]} \|B(t)\| \leq \alpha_B$, $\max_{t\in[0,T]} \|B'(t)\| \leq \beta_B$, $\max_{t\in[0,T]} \|[A, B(t)]\| \leq \alpha_{AB}$ and $\|\cdot\|$ denotes the spectral norm. The oracles themselves carry a non-trivial cost, so the above expression is not the final cost.

For simplicity, we present the detailed computation in Appendix~\ref{B} and summarize the results here. It can be shown that $\max_{t\in[0,T]} \|B(t)\| = N\zeta$, $\max_{t\in[0,T]} \|B'(t)\| = N\zeta\omega$ and 
$\max_{t\in[0,T]} \|[A, B(t)]\| \leq 4JN\zeta d$.

To make a proper comparison, we must also calculate the costs to implement the oracles $O_A$ and $O_B$. In our case, $A$ is diagonal, and so we can treat $O_A$ as diagonal evolution. The cost associated with diagonal evolution is $\mathcal{O}(L\tilde d)$ where $L$ is the number of terms in the diagonal part and $\tilde d$ is the upper bound on the locality of each term~\cite{Kalev_2021}. The summation contains $dN$ terms, hence $L = dN$. Since each term contains 2 $Z$ operators, $\tilde d = 2$. Hence the cost for implementing $O_A$ is $\mathcal{O}(2Nd)\equiv\mathcal{O}(Nd)$.

To calculate the cost of oracle $O_B$, we essentially need to calculate the cost of implementing the operator $B(t)$. The most straightforward implementation uses LCU~\cite{Low_2019b}. As there are $N$ terms in the sum of Pauli-$X$'s, the cost for LCU will be $\mathcal{O}(N^2)$.

Hence, upon simplification, the overall simulation cost scales as
\begin{equation}\label{eq:53}
\begin{aligned}
\mathcal{O}\Bigg(
&\min\Bigg\{
\frac{(N\zeta T)^2}{\epsilon}\,
\log\!\left(\frac{N\zeta T}{\epsilon}\right),
\\
&N\zeta T + \frac{N\zeta\sqrt{Jd+\omega}\,T^{3/2}}{\epsilon^{1/2}}
\\
&\times\log\!\left(\frac{N\zeta T (Jd+\omega)}{\epsilon}\right)
\Bigg\}
\\
&\times
\log\!\left(\frac{N\zeta T(Jd+\omega)}{\epsilon}\right)\times N(d+N)
\Bigg)\ .
\end{aligned}
\end{equation}
The qubit overhead is set by the ancilla registers required for the HAM-$T$ oracles. The ancilla register $a$ uses
$\mathcal{O}(\log N)$ qubits, while the register $m$ uses $\mathcal{O}(\log M)$ qubits. Following Lemmas~8 and~9 of
Ref.~\cite{An_2022}, the number of quadrature nodes satisfies
\begin{equation}
\begin{aligned}
M &= 
    \frac{\alpha_{AB}+\beta_B}{\alpha_B^2}\quad\text{OR}\quad\sqrt{\frac{2(\alpha_{AB}+\beta_B)T}{\epsilon}}\ ,
\end{aligned}
\end{equation}
depending on the choice of the minimum in Eq.~(\ref{eq:47}). Thus the total qubit cost scales as
\begin{equation}
\begin{aligned}
    &\mathcal{O}\left(\log N+\log\left(\frac{Jd+\omega}{N\zeta}\right)\right)\\&\text{OR}\quad\mathcal{O}\left(\log N+\log\left(\frac{(Jd+\omega)T}{\epsilon}\right)\right)\ .
\end{aligned}
\end{equation}
When this expression yields $M < 1$, the constraint is vacuous and we set $M=1$. We can also write this simply as:
\begin{equation}
\begin{aligned}
    &\mathcal{O}\left(\log N+\log\left\lceil\frac{Jd+\omega}{N\zeta}\right\rceil\right)\\&\text{OR}\quad\mathcal{O}\left(\log N+\log\left\lceil\frac{(Jd+\omega)T}{\epsilon}\right\rceil\right)\ .
\end{aligned}
\end{equation}

\subsubsection{With PMR}

We now analyze the same Hamiltonian when simulated using the PMR algorithm. First, the time-dependent part of the Hamiltonian must be expanded in the PMR form. The expansion takes the form:
\begin{equation}\label{eq:54}
V(t) = \sum_{i=1}^M D_i(t) P_i\ ,
\end{equation}
where $M = N$, $P_i = X_i$, and:
\begin{equation}\label{eq:55}
D_i(t) = \sum_{k=1}^K \exp(\Lambda_i^{(k)} t) D_i^{(k)}\ ,
\end{equation}
where $K = 2$, $\Lambda_i^{(k)} = (-1)^k j\omega$ for all $i$ (where $j=\sqrt{-1}$) and $D_i^{(k)} = -\frac{\zeta}{2}I$ for all $i$.

Now we will count each of the parameters as required by the complexity given by Ref.~\cite{Chen_2021}:
\begin{equation}\label{eq:56}
\begin{aligned}
\mathcal{O}(&r(Q^2 + QM(k_{\text{od}} + \log M) \\&+QMK(C_D + C_{\Delta H_0} + C_\Lambda)) + L\tilde d)\ .
\end{aligned}
\end{equation}
We will define each parameter as we calculate it.

$k_{\text{od}}$ is defined as the upper bound on the locality of $P_i$. Since each $P_i$ matrix is a tensor product including at most a single $X$ operator, $k_{\text{od}} = 1$.

Next, $r$ is defined as the total number of small time evolutions. In turn, each small time evolution can have a different size depending on the off-diagonal norm. $\lambda := \max_{i,k,z}\{\Re(\langle z|\Lambda_i^{(k)}|z\rangle)\}$ is defined as the maximum of the real part of all eigenvalues from all the matrices $\Lambda_i^{(k)}$. According to Ref.~\cite{Chen_2021}, when this is 0, the total time $T$ is partitioned into intervals defined by $\Delta t_w = \ln 2/\Gamma(t_w)$. So the time interval depends on the off-diagonal norm at that time. We can try to find an upper bound on $r$ by finding a lower bound on $\Delta t_w$:
\begin{equation}\label{eq:57}
\Delta t_w = \frac{\ln 2}{\Gamma(t_w)} = \frac{\ln 2}{N\zeta|\cos(\omega t_w)|} \geq \frac{\ln 2}{N\zeta}\ .
\end{equation}
Hence, the upper bound on $r$ will be $r \leq TN\zeta/\ln 2$.

$L$ is defined as the number of terms in the time-independent Hamiltonian, which is clearly $dN$, and $\tilde d$ is defined as the upper bound on locality of each of the terms. Since each term contains 2 $Z_i$ operators, $\tilde d = 2$.

$Q$ is the truncation order to be used in the LCU routine of the circuit implementation. This has the same dependence as seen in the time-independent case, i.e.,
\begin{equation}\label{eq:58}
Q = \mathcal{O}\left(\frac{\log(r/\epsilon)}{\log\log(r/\epsilon)}\right)\ .
\end{equation}
As we saw earlier, we can omit this from the final cost up to polylogarithmic corrections because of its asymptotic behavior.

$C_D$ and $C_\Lambda$ are the costs of obtaining any matrix element of $D_i^{(k)}$ and $\Lambda_i^{(k)}$ respectively. Since $D_i^{(k)} \propto I$, we have $C_D = \mathcal{O}(1)$. Similarly, since $\Lambda_i^{(k)}$ are 1-dimensional, $C_\Lambda = \mathcal{O}(1)$. Similar to the time-independent case, $C_{\Delta H_0}$ is the cost of calculating the change to a diagonal energy due to the action of a $P_i$. Since $P_i=X_i$, acting a permutation on $H_0$ changes the sign of $2d$ terms from the sum. Thus the cost is $\mathcal{O}(d)$.

Hence upon simplification, the final gate cost becomes:
\begin{equation}\label{eq:59}
\begin{aligned}
&\mathcal{O}(N\zeta T(Q^2 + QN\log N + QNd) + Nd) \\
&\approx \mathcal{O}(N^2\zeta T(\log N + d))\ .
\end{aligned}
\end{equation}
In addition, following the discussion in Ref.~\cite{Chen_2021}, the qubit cost arises from the state preparation in the LCU procedure, costing $\mathcal{O}(Q\log(MK))\approx\mathcal{O}(\log (N))$ ancilla qubits.

\subsubsection{Comparison}

\begin{table*}[tb]
\begin{center}
\begin{tabular}{| c | c | c |} 
 \hline \hline
 Algorithm & Gate Cost & Qubit Cost\\ [0.5ex]
 \hline
 qHOP & $\mathcal{O}\Bigg(\min\Bigg\{\frac{(N\zeta T)^2}{\epsilon}\log\left(\frac{N\zeta T}{\epsilon}\right),N\zeta T+\frac{N\zeta(Jd+\omega)^{1/2}T^{3/2}}{\epsilon^{1/2}}\log\left(\frac{N\zeta T(Jd+\omega)}{\epsilon}\right)\Bigg\}$ & $\mathcal{O}\left(\log N+\log\left\lceil\frac{Jd+\omega}{N\zeta}\right\rceil\right)$\\
 & $\times\log\left(\frac{N\zeta T(Jd+\omega)}{\epsilon}\right)\times N(d+N)\Bigg)$ & $\text{OR}\quad\mathcal{O}\left(\log N+\log\left\lceil\frac{(Jd+\omega)T}{\epsilon}\right\rceil\right)$\\ [2ex]
 \hline
 PMR & $\mathcal{O}(N^2\zeta T(\log N+d))$ & $\mathcal{O}(\log N)$ \\ [1ex]
 \hline
\end{tabular}
\caption{\label{tab:timedep}\textbf{Comparison of algorithm costs for time-dependent simulation.} While with qHOP we can see that for high values of $\omega$, the minimum would be independent of $\omega$, thus making it useful for highly oscillatory dynamics. However, the PMR approach completely gets rid of dependence on $\omega$, making it useful in all regimes. The qubit cost for the PMR approach depends only on the system size, whereas the qubit cost for qHOP depends on other system parameters.
}
\end{center}
\end{table*}

The two algorithms exhibit distinct scaling characteristics, as summarized in Table~\ref{tab:timedep}. 

The scaling with $N$ differs between the approaches. PMR exhibits $\mathcal{O}(N^2\log N)$ scaling, which may be favorable for simulations involving a large number of qubits compared to qHOP's higher polynomial scaling in $N$ (for both choices in the minimum). This difference arises from the fact that qHOP employs block encoding and QSP for both $A$ and $B(t)$, with both contributions dependent on $N$, whereas PMR treats the diagonal part $A$ as energy shifts that appear in the expansion as phases, reducing one source of $N$ dependence.

Another noteworthy distinction is that PMR's cost is completely independent of the driving frequency $\omega$, while qHOP's cost has a logarithmic dependence on $\omega$. In PMR, the phases in Eq.~(\ref{eq:32}) are integrated out into the expansion~(\ref{eq:35}) using the identity~(\ref{eq:34}), and the bound of each term is independent of the oscillations. In contrast, qHOP's bounds explicitly depend on the rate of change $B'(t)$, which scales with $\omega$. Note, however, that for large $\omega$, the chosen minimum will be independent of $\omega$, showcasing the benefit of qHOP for highly oscillatory dynamics.

The scaling with spatial dimension $d$ also differs. PMR exhibits linear scaling with $d$, while qHOP shows $\mathcal{O}(d\log d)$ dependence in the best case. The parameter $d$ appears in the context of the diagonal time-independent term, with $dN$ total terms in the summation. In the interaction picture, PMR treats diagonal evolution as energy shifts that appear as phases in the expansion. Thus, the complexity depends primarily on the time-dependent part, with diagonal evolution contributing only through the $\mathcal{O}(L\tilde d)$ term when transforming back to the Schr\"odinger picture. Along with the above, the change in the diagonal energy by a single permutation also just involves $\mathcal{O}(d)$ terms. For qHOP, one of the bounds involves the time-independent term, leading to $d$ dependence in the oracle query complexity, with additional contributions from implementing $O_A$.

Similarly, PMR's cost does not depend on the interaction energy $J$, while qHOP has explicit $J$ dependence. This follows from PMR's treatment of the diagonal part as energy shifts in the series expansion of $U$, which affects the cost only when transforming back to the Schr\"odinger picture.

When comparing qubit requirements, the two methods exhibit distinct parameter dependencies. The qubit cost of qHOP depends on physical system parameters rather than system size, whereas the qubit cost of PMR is governed primarily by the system size. In both cases, however, the qubit overhead scales logarithmically with the relevant parameters.

\section{Summary and conclusion}\label{sec:summary}

In this work, we presented a comparative study of the permutation matrix representation (PMR) algorithm for Hamiltonian simulation, examining its performance relative to state-of-the-art methods such as qubitization and qHOP. Our analysis, applied to representative time-independent and time-dependent models, highlights structural advantages of PMR that are particularly relevant in the resource-limited era of near-term quantum hardware.

A central feature of PMR is its separation of diagonal and off-diagonal contributions to the Hamiltonian. This decomposition allows diagonal terms to be treated as phase evolutions rather than explicit gate operations, effectively removing their contribution to circuit depth and ancilla overhead. Since many current platforms---including superconducting qubits, trapped ions, and Rydberg arrays---natively implement diagonal operations at low cost, this property aligns PMR naturally with existing hardware capabilities. Off-diagonal components, represented as permutation operators, can likewise often be realized using native bit-flip or spin-exchange gates. In contrast, qubitization and qHOP rely on block-encoding constructions and quantum signal processing, which require deeper circuits, multiple ancilla registers, and complex reflections about entangled ancilla states. While these methods achieve optimal asymptotic performance, their practical implementation on near-term devices is constrained by coherence-time limitations, restricted connectivity, and the absence of robust mid-circuit control.

The time-dependent formulation of PMR exhibits analogous advantages. By expressing the Dyson series through divided differences, PMR avoids explicit quadrature, oracle calls to evaluate the Hamiltonian at multiple time slices, and dependencies on derivatives or commutators. Because PMR integrates temporal effects directly into its expansion coefficients, the algorithm's cost remains stable even in regimes involving rapid modulation or strong periodic driving---a property especially relevant for platforms employing Floquet engineering and fast parametric modulation.

Our case studies illustrate how these structural features translate into concrete resource benefits. For time-independent Rydberg Hamiltonians, PMR avoids the quadratic growth in block-encoding complexity associated with long-range interactions, while maintaining comparable qubit overhead. For the Floquet-driven transverse-field Ising model, PMR does not incur explicit gate-count scaling with the driving frequency, interaction strength, or spatial dimensionality in the manner required by qHOP. Taken together, these analyses suggest that PMR offers a compelling alternative to QSP-based methods in regimes where diagonal terms dominate, off-diagonal operators are sparse, or time dependence is strong.

While PMR does not supplant qubitization or QSP in the long-term landscape of fault-tolerant quantum computation, it occupies a particularly promising niche in the pre-fault-tolerant era. Its low ancilla requirements, modest circuit depth, compatibility with analog primitives, and resilience to rapid temporal variation make it well suited for near-term demonstrations of quantum simulation. As hardware evolves toward hybrid analog--digital architectures, the structure of PMR aligns naturally with emerging device capabilities and may enable early instances of quantum advantage in dynamical simulation. Future work may focus on noise-aware variants of PMR, hardware-native implementations, and hybrid approaches that combine the structural simplicity of PMR with the optimal asymptotic scaling of QSP-based methods.
\\

\begin{acknowledgments}
This research was sponsored by the Secretary of the Air Force Concepts, Development, and Management (SAF/CDM) organization through the SEQCURE2 program at the University of Maryland's Applied Research Laboratory for Intelligence and Security. H.S. acknowledges support from the Graduate School Fellowship awarded by the USC Dana and David Dornsife College of Letters, Arts and Sciences.
\end{acknowledgments}

\bibliography{refs}

\appendix

\section{LCU block encoding for the Rydberg Hamiltonian}\label{A}
Recall the Hamiltonian~(\ref{eq:rh}). Let us define $C_{ij}:=\frac{C_6}{|\textbf{r}_i-\textbf{r}_j|^6}$. First let us compute the interaction terms in terms of Pauli strings:
\begin{equation}
\resizebox{\columnwidth}{!}{$
\begin{aligned}
\sum_{i<j}C_{ij}n_in_j&=\sum_{i<j}C_{ij}(I+Z_i)(I+Z_j)\\&=\sum_{i<j}C_{ij}I+\sum_{i<j}C_{ij}Z_i+\sum_{i<j}C_{ij}Z_j+\sum_{i<j}C_{ij}Z_iZ_j\\&=C\cdot I+\sum_{i=1}^{N-1}C'_iZ_i+\sum_{j=2}^NC''_jZ_j+\sum_{i<j}C_{ij}Z_iZ_j\ ,
\end{aligned}
$}
\end{equation}
where we have converted the double sums to single sums by defining $C'_i$ and $C''_j$, which represent $C_{ij}$ summed over $j$ and $i$ respectively (following the condition $i<j$ throughout). Here $C'_N=C''_1=0$ because of the condition $i<j$. The total Hamiltonian can then be written as:
\begin{equation}
\resizebox{\columnwidth}{!}{$
\begin{aligned}
H_{\text{Ryd}}&=C\cdot I+\frac{1}{2}\sum_{i=1}^N\Omega_iX_i+\sum_{i=1}^N(C'_i+C''_i-\frac{\delta}{2})Z_i+\sum_{i<j}C_{ij}Z_iZ_j\\&\equiv\frac{1}{2}\sum_{i=1}^N\Omega_iX_i+\sum_{i=1}^N(C'_i+C''_i-\frac{\delta}{2})Z_i+\sum_{i<j}C_{ij}Z_iZ_j\\&:=\sum_{i=1}^M\alpha_iP_i\ ,
\end{aligned}
$}
\end{equation}
where $\alpha_i\geq0$ and $P_i$ are Pauli strings which include any negative signs. We have also omitted the Identity term as it amounts to an overall phase, and $M=\mathcal{O}(N^2)$ because of the $\sum_{i<j}$ term.

\section{Upper bounds for the qHOP algorithm}\label{B}
Let us calculate the upper bounds in Eq.~(\ref{eq:47}) for the Hamiltonian~(\ref{eq:46}). For the interaction picture analysis, the Hamiltonian can be decomposed into $A$ and $B(t)$ as:
\begin{equation}
    A=-J\left(\sum_{\langle i,j\rangle}Z_iZ_j\right)
\end{equation}
\begin{equation}
    B(t)=-\zeta\cos(\omega t)\left(\sum_iX_i\right)\ .
\end{equation}
Firstly, note that $B(t)$ is diagonal in the Hadamard basis $|{\pm}\rangle^{\otimes N}=|{\pm}\rangle\otimes\cdots\otimes|{\pm}\rangle$ where $|{\pm}\rangle=\frac{1}{\sqrt{2}}(|0\rangle\pm|1\rangle)$. Since $\sum_iX_i$ is Hermitian, the spectral norm equals its largest absolute eigenvalue. The largest eigenvalue follows immediately:
\begin{equation}
    \sum_{i=1}^NX_i|{+}\rangle^{\otimes N}=N|{+}\rangle^{\otimes N}\ .
\end{equation}
Since $|\cos(\cdot)|\leq 1$ and $|\sin(\cdot)|\leq 1$, the maxima over the total time are:
\begin{equation}
    \max_{t\in[0,T]}\norm{B(t)}=\zeta\ \norm{\sum_{i=1}^NX_i}=N\zeta 
\end{equation}
\begin{equation}
    \max_{t\in[0,T]}\norm{B'(t)}=\zeta\omega\ \norm{\sum_{i=1}^NX_i}=N\zeta\omega\ .
\end{equation}
For the remaining upper bound, let us first explicitly calculate the commutator:
\begin{equation}
\begin{aligned}
    \left[A,B(t)\right]&=J\zeta\cos(\omega t)\left[\sum_{\langle i,j\rangle}Z_iZ_j\ ,\sum_iX_i\right]\\&=2iJ\zeta\cos(\omega t)\left(\sum_{\langle i,j\rangle}Y_iZ_j\ +\sum_{\langle i,j\rangle}Z_iY_j\right)\ .
\end{aligned}
\end{equation}
We can then use the triangle inequality to show:
\begin{equation}
\resizebox{\columnwidth}{!}{$
\begin{aligned}
    \max_{t\in[0,T]}\norm{\ [A,B(t)]\ }&\leq2J\zeta\left(\norm{\sum_{\langle i,j\rangle}Y_iZ_j}+\norm{\sum_{\langle i,j\rangle}Z_iY_j}\right)\\&\leq2J\zeta\left(\sum_{\langle i,j\rangle}\norm{Y_iZ_j}+\sum_{\langle i,j\rangle}\norm{Z_iY_j}\right)\\&=4JN\zeta d\ .
\end{aligned}
$}
\end{equation}

\section{$C_{D_0}$ approximation}\label{C}

We want to cut off the second term in $D_0$ such that there is just a small error in implementing $e^{-iD_0\Delta t}$. We can do this using the following method.

For two diagonal operators $A$ and $B$, if $\norm{A-B}\leq\epsilon$, then $\norm{e^{-i\theta A}-e^{-i\theta B}}\leq|\theta|\epsilon$. Let $\tilde{D}_0$ be the approximation of $D_0$ after the cutoff. Then $\norm{D_0-\tilde{D}_0}\leq\epsilon$ implies $\norm{e^{-i\Delta tD_0}-e^{-i\Delta t\tilde{D}_0}}\leq\epsilon\Delta t$. Since the diagonal evolutions are repeated $r$ times, the error for each short time evolution must satisfy $\leq \epsilon/r$. Hence, to ensure $\norm{e^{-i\Delta tD_0}-e^{-i\Delta t\tilde{D}_0}}\leq\frac{\epsilon}{r}$, it suffices that $\norm{D_0-\tilde{D}_0}\leq\frac{\epsilon}{t}$.

The aim now is to get a condition on the cutoff $n$ such that:
\begin{equation}
    C_6'\norm{\sum_{i<j}\frac{1}{|i-j|^6}Z_iZ_j-\sum_{\substack{i<j\\|i-j|\leq n}}\frac{1}{|i-j|^6}Z_iZ_j}\leq\frac{\epsilon}{t}\ .
\end{equation}
The term inside $\norm{...}$ is a diagonal matrix, with maximum possible entry $\sum_{\substack{i<j\\|i-j|>n}}\frac{1}{|i-j|^6}$. Now $\forall \ i,j$, the maximum possible value for this is $\sum_{i=n+1}^{N-1}i^{-6}$. The inequality then boils down to showing:
\begin{equation}
    \sum_{i=n+1}^{N-1}\frac{1}{i^6}\leq\frac{\epsilon}{tC_6'}\ .
\end{equation}
It suffices to show this for $N\rightarrow \infty$. We can use the following inequality:
\begin{equation}
    \sum_{i=n+1}^{\infty}\frac{1}{i^6}\leq\int_n^\infty\frac{\mathrm{d}x}{x^6}=\frac{1}{5n^5}\ .
\end{equation}
Thus it suffices to show 
\begin{equation}
    \frac{1}{5n^5}\leq\frac{\epsilon}{tC_6'}\ .
\end{equation}
This is satisfied for 
\begin{equation}
    n\geq\left(\frac{tC_6'}{5\epsilon}\right)^{1/5}\ .
\end{equation}
Thus the minimum cutoff is $\mathcal{O}\left(\left(\frac{tC_6'}{\epsilon}\right)^{1/5}\right)$. Hence $C_{D_0}=\mathcal{O}\left(N\left(\frac{tC_6'}{\epsilon}\right)^{1/5}\right)$.

\section{$C_{\Delta D_0}$ approximation}\label{D}

We want to apply essentially the same method to $C_{\Delta D_0}$. Suppose we act a permutation $P_k\ (=X_k)$ to $D_0$ as $X_kD_0X_k$, then $C_{\Delta D_0}$ is determined by how many terms change their sign. The difference in the diagonal energies is determined by:
\begin{equation}
\begin{aligned}
    \Delta D_0&:=D_0-X_kD_0X_k\\&=Z_k\left(-\delta +2\sum_{\substack{i=1\\i\neq k}}^N\frac{C_6'}{|k-i|^6}+2\sum_{\substack{i=1\\i\neq k}}^N\frac{C_6'}{|i-k|^6}Z_i\right)\\&:=Z_k\left(-\delta+\#+2\sum_{\substack{i=1\\i\neq k}}^N\frac{C_6'}{|i-k|^6}Z_i\right)\ ,
\end{aligned}
\end{equation}
where $\#$ is some number. The last sum has $N$ terms, implying that $C_{\Delta D_0}=\mathcal{O}(N)$. However, by truncating the sum, we obtain an approximation of $\tilde U_{\text{od}}$ that can be implemented with bounded error.

First, we can use the Hermite-Genocchi identity to show that if $|x_i-\tilde{x}_i|\leq\epsilon$ for some set of numbers $\{x_i\}_{i=0}^q$ and $\{\tilde{x}_i\}_{i=0}^q$, then:
\begin{equation}
    \left|e^{-i\Delta t[\tilde x_0,...,\tilde x_q]}-e^{-i\Delta t[x_0,...,x_q]}\right|\leq\frac{\Delta t^{q+1}\epsilon}{q!}+\mathcal{O}(\epsilon^2)\ ,
\end{equation}
which implies
\begin{equation}
    \Rightarrow\left|\frac{e^{-i\Delta t[\tilde x_0,...,\tilde x_q]}}{\Delta t^q/q!}-\frac{e^{-i\Delta t[x_0,...,x_q]}}{\Delta t^q/q!}\right|\leq\epsilon\Delta t+\mathcal{O}(\epsilon^2)\ .
\end{equation}
This means that $|\beta_{\mathbf{i}_q}^{(z)}-\tilde\beta_{\mathbf{i}_q}^{(z)}|\leq\left|\frac{d_{\mathbf{i}_q}}{\Gamma_{\mathbf{i}_q}}\right|\epsilon\Delta t\leq\epsilon\Delta t$. We can write $\tilde U_{\text{od}}$ as:
\begin{equation}
\begin{aligned}
    \tilde U_{\text{od}}&=\sum_z\sum_{q=0}^Q\sum_{\mathbf{i}_q}\frac{(\Delta t)^q}{q!}\Gamma_{\mathbf{i}_q}\beta_{\mathbf{i}_q}^{(z)}P_{\mathbf{i}_q}\ket{z}\bra{z}\\&=\sum_{q=0}^Q\sum_{\mathbf{i}_q}\frac{(\Delta t)^q}{q!}\Gamma_{\mathbf{i}_q}P_{\mathbf{i}_q}\left(\sum_z\beta_{\mathbf{i}_q}^{(z)}\ket{z}\bra{z}\right)\ .
\end{aligned}
\end{equation}
Thus we can write the error in $\tilde U_{\text{od}}$ as:
\begin{equation}
\begin{aligned}
    &\norm{\tilde U_{\text{od}}-\tilde{\tilde U}_{\text{od}}}\\&=\norm{\sum_{q=0}^Q\sum_{\mathbf{i}_q}\frac{\Delta t^q}{q!}\Gamma_{\mathbf{i}_q}P_{\mathbf{i}_q}\left(\sum_z\left(\beta_{\mathbf{i}_q}^{(z)}-\tilde \beta_{\mathbf{i}_q}^{(z)}\right)\ket{z}\bra{z}\right)}\\&\leq\sum_{q=0}^Q\sum_{\mathbf{i}_q}\frac{\Delta t^q}{q!}\Gamma_{\mathbf{i}_q}\norm{P_{\mathbf{i}_q}}\norm{\sum_z\left(\beta_{\mathbf{i}_q}^{(z)}-\tilde \beta_{\mathbf{i}_q}^{(z)}\right)\ket{z}\bra{z}}\\&\leq\epsilon\Delta t\sum_{q=0}^Q\sum_{\mathbf{i}_q}\frac{\Delta t^q}{q!}\Gamma_{\mathbf{i}_q}\approx2\epsilon\Delta t\ ,
\end{aligned}
\end{equation}
because $\norm{P_{\mathbf{i}_q}}=1$. So to get an error $\epsilon/r$ in implementing $\tilde U_{\text{od}}$, it suffices to show $|x_i-\tilde x_i|\leq\epsilon/2t$.

Now that we know how the error propagates, let us compute the approximation. Let $\Delta\tilde D_0$ be the approximation of $\Delta D_0$ defined as:
\begin{equation}
    \Delta\tilde D_0:=Z_k\left(-\delta+\#+2\sum_{\substack{i=1\\i\neq k\\|i-k|\leq n}}^N\frac{C_6'}{|i-k|^6}Z_i\right)\ .
\end{equation}
If we are able to restrict $\norm{\Delta\tilde D_0-\Delta D_0}\leq\epsilon/2t$, the error in $\Delta E$'s (the eigenvalues) will also be $\leq \epsilon/2t$. However, there is still a small caveat. The calculation for $\Delta D_0$ is only for the action of a single permutation operator, whereas the change in energy $\Delta E_{z_q}$ requires the action of $q$ permutations. The maximum value of $q$ is $Q$, hence we will have to bound the error for a single permutation by $\epsilon/2Qt$ so that the total error for $\Delta E_{z_Q}$ does not stack up more than $\epsilon/2t$.

The task now is to find a condition on $n$ such that:
\begin{equation}
    2C_6'\norm{\sum_{\substack{i=1\\i\neq k}}^N\frac{1}{|i-k|^6}Z_i-\sum_{\substack{i=1\\i\neq k\\|i-k|\leq n}}^N\frac{1}{|i-k|^6}Z_i}\leq\epsilon/2Qt\ .
\end{equation}
Proceeding similar to $C_{D_0}$, the matrix inside the norm is a diagonal matrix, with the maximum possible entry being $\sum_{\substack{i=1\\|i-k|>n}}^N\frac{1}{|i-k|^6}$. This number is smaller than $2 \sum_{i=n+1}^\infty i^{-6}$. We thus have to find an $n$ such that:
\begin{equation}
    2 \sum_{i=n+1}^\infty\frac{1}{i^6}\leq \epsilon/4QtC_6'\ .
\end{equation}
Using the same integral inequality as the $C_{D_0}$ case, we get
\begin{equation}
    n\geq\left(\frac{8QtC_6'}{5\epsilon}\right)^{1/5}\ ,
\end{equation}
which implies that the minimum cutoff is $\mathcal{O}\left(\left(\frac{QtC_6'}{\epsilon}\right)^{1/5}\right)$. Hence $C_{\Delta D_0}=\mathcal{O}\left(\left(\frac{tC_6'}{\epsilon}\right)^{1/5}\right)$, where we have omitted the $\log/\log\log$ factor $Q$.

\end{document}